\documentclass{aa}  

\usepackage{txfonts}
\usepackage{lipsum}
\usepackage{subcaption}         
\usepackage{lscape}             
\usepackage{placeins}           
                                
\usepackage{graphicx}	
\usepackage{amsmath}	
\usepackage{booktabs} 
\usepackage{xcolor}
\usepackage{hyperref}

\def\Planck{\textit{Planck}}

\begin{document}
\title{ComPACT: Mass–Redshift Properties of the galaxy cluster catalogue}

\author{S. Voskresenskaia\inst{1,2}\fnmsep\thanks{E-mail: voskr@cosmos.ru} \and 
N. Lyskova\inst{1} \and
I. Zaznobin\inst{1} \and
A. Meshcheryakov\inst{1}\thanks{E-mail: mesch@cosmos.ru}}

\institute{
Space Research Institute (IKI) Russian Academy of Sciences, Profsoyuznaya 84/32, Moscow 117997, Russia
\and
HSE University, 20 Myasnitskaya St., Moscow 101000, Russia}

\date{Accepted XXX. Received YYY; in original form ZZZ}



\abstract
{Machine-learning methods are increasingly applied to astronomical surveys, providing powerful tools for detecting and studying galaxy clusters.}
{We investigate the mass-redshift properties and completeness of the \textsc{ComPACT} galaxy cluster catalogue, constructed using a convolutional neural network applied to  publicly available combined ACT+\Planck\ maps.}
{The ComPACT catalogue contains 2,962 SZ-selected galaxy cluster candidates. We confirm clusters by estimating redshifts using literature information and photometric techniques based on DESI Legacy Imaging Surveys data. Cluster masses are derived from ACT+\Planck\ and \Planck\ Compton-y maps via SZ scaling relations. The completeness is assessed using simulated cluster injections into real microwave maps.}
{We confirm approximately $\sim$60\% of the ComPACT candidates as galaxy clusters. The redshifts span the range $0.007 < z < 1.7$, including approximately 116 new measurements.  Masses are obtained for 56\% of the sample, covering the range $(0.25 - 13.1) \times 10^{14} ~M_\odot$ and including 158 new mass determinations. We identify five previously unreported massive clusters ($M_{500c} > 6 \times 10^{14}~M_\odot$) at $z > 0.7$, increasing the known population of such systems by approximately 10 \%.}
{The ComPACT catalogue expands the SZ-selected \Planck-like cluster population, especially at high redshift and high mass, demonstrating the effectiveness of deep-learning approaches for cluster detection in microwave data.}


\maketitle
\nolinenumbers

\section{Introduction}
\label{sec:intro}
\thispagestyle{empty}

Galaxy clusters are the most massive gravitationally bound systems in the Universe and constitute key probes of cosmology and galaxy evolution. They form at the sites of rare, high-amplitude primordial matter density fluctuations in the early Universe \citep{Sarazin_1988}. As the largest virialised structures, galaxy clusters provide valuable laboratories for studying the growth of large-scale structure, cosmic evolution, and the interplay between dark matter, the hot intracluster medium, and cluster galaxies \citep[see, e.g.,][for reviews]{Clusters_Allen,2012ARA&A..50..353K}. The total mass of galaxy clusters is dominated by dark matter (approximately 85\%), which defines the gravitational potential well that confines the hot, X-ray–emitting intracluster gas.

Photons undergo inverse Compton scattering off the hot electrons in galaxy clusters, resulting in a characteristic distortion of the cosmic microwave background blackbody spectrum known as the thermal Sunyaev–Zeldovich (tSZ) effect \citep{1970Ap&SS...7....3S,1972CoASP...4..173S}. Owing to the redshift independence of the tSZ surface brightness, galaxy clusters can be detected in the microwave band out to the highest redshifts. 
The first galaxy clusters discovered in a blind survey via their tSZ signature were reported more than a decade ago \citep{Staniszewski_2009}. Since then, substantial progress in tSZ-selected cluster studies has been achieved by the South Pole Telescope (SPT; \citealt{SPT100,SPTpol,SPT_MCMF}), the Atacama Cosmology Telescope (ACT; \citealt{Hilton_2018,Hilton_2021,ACT-MCMF,ACT2025}), and the \Planck\ satellite mission \citep{Planck_2014,Planck_2016}. 

Cluster redshifts are generally determined from the spectroscopy or photometry of their member galaxies \citep[e.g.][]{SPT2500D}, providing distance estimates that are essential for mapping large-scale structure and for cosmological analyses. Cluster masses, in turn, can be inferred using a variety of  techniques that probe different cluster components \citep[see, e.g.][for a review]{Pratt_2019SSRv..215...25P, 2025arXiv250507697M}. A number of X-ray mass proxies have been proposed  and are commonly used to determine  individual cluster masses \citep[][among others]{2006ApJ...650..128K, 2009A&A...498..361P, Arnaud_2010, 2018MNRAS.473.3072M, 2015MNRAS.450.1984C, 2022A&A...665A..24P, 2025A&A...702A.175L, 2025JCAP...05..007K}. The total SZ flux also correlates tightly with the total cluster mass \cite[e.g.,][among many others]{2006ApJ...650..538N, 2010A&A...517A..92A, Planck_2014}. Weak gravitational lensing enables a direct probe of the underlying dark matter potential \citep[e.g.][]{2019MNRAS.482.1352M}, independent of the cluster dynamical state. Dynamical analysis based on the galaxy velocity dispersions, on par with weak lensing techniques, is considered to provide unbiased cluster mass estimates \citep[e.g.][]{2008ApJ...672..122E, 2013ApJ...772...47S, 2021A&A...655A.115F}. All these different techniques establish the connection between observed quantities and the underlying halo properties, enabling the use of galaxy clusters as cosmological probes and as environments for studying baryonic processes in dense regions.

Over the past decade, several extensive catalogues of SZ-selected galaxy clusters have been compiled. The full-sky survey conducted by \Planck\ resulted in the PSZ2 catalogue \citep{Planck_2014}, which includes 1,653 SZ-detected objects, 1,334 of which have been confirmed through optical observations \citep{UPCluster-SZ}. The Atacama Cosmology Telescope (ACT), in its fifth data release (DR5), provided over 4,000 optically confirmed SZ clusters across an area of approximately 13,211 deg$^2$ \citep{Hilton_2021}. More recently, the ACT DR6 release reported around 10,000 galaxy clusters \citep{ACT2025}, significantly surpassing the size of previous SZ catalogues. The South Pole Telescope (SPT) identified 677 cluster candidates in a 2,500 deg$^2$ survey, of which 516 have been confirmed \citep{SPT-SZ}. These catalogues are further complemented by more recent datasets from SPTpol \citep{SPTpol} and SPT-DEEP \citep{2025arXiv250317271K}. 

To increase the number of galaxy clusters identified in current data, several complementary strategies have been pursued. One approach involves combining microwave observations from multiple instruments, as demonstrated by the joint analysis of \Planck\ and SPT data (PSZSPT; \citealt{PSZSPT}), which revealed several dozen clusters absent from the individual SPT/\Planck\ catalogues. Another method exploits the synergy between different wavebands, for example, by cross-correlating microwave and X-ray observations. \citet{ComPRASS} combined \Planck\ and ROSAT all-sky survey data to construct the ComPRASS catalogue, which contains nearly 2000 joint X-ray–SZ detections, around one quarter of which were previously unknown. A further avenue involves extending existing cluster-confirmation algorithms. For instance, \citet{ACT-MCMF, SPT_MCMF, PSZ_MCMF} applied the multi-component matched filter approach with optical survey data to follow up low signal-to-noise (SNR) SZ detections, enabling the discovery of a large number of lower-mass systems not identified in published catalogues.

During the past decade, machine learning (ML) techniques have become increasingly prominent in the analysis of tSZ data. These methods enable data-driven inference, allowing signal detection and classification without relying on predefined models or templates \citep[for a  review, see][]{ML_review_2023RPPh...86g6901M}. In the context of SZ studies, deep learning approaches have proven particularly effective for signal extraction and component separation. For example, \citet{Bonjean_2020} and \citet{SZcat} applied a U-Net convolutional neural network (CNN) to the multifrequency, full-sky \Planck\ maps. \citet{SZcat} network successfully recovered all previously known \Planck\ SZ clusters and identified a large number of new cluster candidates through component separation. Similarly, \citet{Lin2021} trained a CNN on simulated microwave intensity maps and compared its performance to the conventional matched filter (MF) approach. Although both methods recovered high-SNR systems, the CNN was able to identify lower-SNR clusters missed by the MF. Moreover, combining the outputs of both methods improved overall completeness at fixed purity. In the optical domain, \citet{2023A&A...677A.101G, 2025A&A...695A.246G} developed the YOLO--CL algorithm, an object detection CNN trained on SDSS colour images using redMaPPer clusters as training labels. The model successfully recovered 95--98\% of known clusters with comparable purity, illustrating the effectiveness of modern deep learning techniques in identifying cluster-like structures across multiple wavelengths. 

In this work, we analyse an SZ galaxy cluster catalogue based on machine learning, presented in \cite{Voskresenskaia_2024}. The catalogue, referred to as ComPACT, was constructed using a deep learning, CNN-based method, applied to combined ACT+\Planck\ intensity maps. It was built by targeting regions surrounding SZcat sources \citep{SZcat} with the aim of identifying \Planck-like clusters that fall below the detection threshold of the original \Planck\ survey. A key advantage of this deep-learning approach over traditional matched-filter techniques, as demonstrated in \cite{Voskresenskaia_2024}, is its ability to learn complex, non-linear patterns in the data directly, potentially leading to improved sensitivity to lower signal-to-noise systems compared to methods that rely on a fixed cluster template (e.g. \citealt{Lin2021, Bonjean_2020, Bonjean_2024, Ntampaka_2019}). 

The primary goal of this paper is twofold. First, we  assess the quality and robustness of the ComPACT catalogue by evaluating its completeness and identifying potential biases using simulations. Second, we investigate the redshifts and masses of its constituent clusters in order to characterise the population uncovered by the DL detection method.    Redshifts are obtained using a combination of photometric techniques and cross-matching with existing spectroscopic and photometric redshift data, while masses are estimated via $Y$–$M$ scaling relations from SZ observables measured in this work.

This paper is organised as follows. In Section~\ref{sec:data}, we describe the data sources and external catalogues used in our analysis. Section~\ref{sec:redshift} outlines the methodology for photometric redshift estimation, while Section~\ref{subsection:mass} details the procedures for deriving cluster masses. Section~\ref{sec:results} presents  main results, including completeness and empirical purity, as well as a discussion of particularly notable cluster candidates. We conclude with a summary of our findings in Section~\ref{sec:conclusions}. Masses are reported in terms of $M_{500c}$ if not stated otherwise, where $M_{500c}$ is defined as the mass enclosed within a radius $R_{500c}$ at which the average density is 500 times the critical density at the cluster redshift. We assume a flat $\Lambda{}CDM$ cosmology with $\Omega_m$ = 0.3, $\Omega_\Lambda$ = 0.7, and $H_0$ = 70 km s$^{-1}$ Mpc$^{-1}$ throughout.

\section{Data}
\label{sec:data}
\thispagestyle{empty}
\begin{table*}[ht!]
    \centering
    \begin{tabular}{l l l l l l l} 
        \toprule
        External catalogue & Footprint  & Total No.  & Cross-matched  & Add $z$  & Add $M_{500c}$  & Reference \\
                           & ($deg^2$) & of clusters & clusters &  & by priority & \\
        \midrule
        Up-Cluster-SZ      & full    & 1653  &  466 & 48 & 453  & \cite{UPCluster-SZ} \\ 
        PSZ1               & full    & 1227  &  466 & 24  &      & \cite{PSZ1} \\ 
        PSZ1 subsample     & full    & 115   &  4   & 4   &      & \cite{PSZ1_spec} \\      
        PSZ-MCMF           & 5000    &  853  &  363 & 40 & 169  & \cite{PSZMCMF} \\
        ACT DR6            & 16293   & 9977  & 1362 & 472 & 636  & \cite{ACT2025}  \\ 
        ACT DR5            & 13211   & 4195  & 1027 & 209   & 9    & \cite{Hilton_2021}  \\ 
        ACT-MCMF           & 13211   & 6237  & 1142 & 360  & 16   & \cite{ACT-MCMF}     \\ 
        SPT-DEEP            & 100     &  546  &  13  & 4   & 2    & \cite{2025arXiv250317271K} \\
        SPT2500D           & 2500    &  677  & 163  & 22  & 1    & \cite{SPT2500D} \\
        SPT-ECS            & 2500    &  483  & 150  & 12   & 1    & \cite{SPTpol}       \\ 
        SPT-MCMF           & 2500    &  811  & 161  & 19   & 1    & \cite{SPTMCMF} \\
        \midrule
        ComPRASS           & {full}  & 2323  &  644 & 6   & 103  & \cite{ComPRASS}     \\ 
        \midrule
        MCXC II            & {full}  & 2221  &  326 & 71  & 13   & \cite{MCXCII}       \\
        RASS-MCMF          & 25000   & 8449  &  745 & 98  & 23   & \cite{RASS_MCMF} \\
        MARD-Y3            & 5000    & 2959  &  316 & 13  &      & \cite{MARD-Y3}      \\
        XCSDR1             & 20000   &  503  &   11 & 1   &      & \cite{XCSDR1}       \\ 
        CODEX              & 10382   & 10382 &  331 & 16  &      & \cite{CoDEX}        \\
        \midrule
        Wen \& Han High-z  & 10000   & 1959  &   21 & 6   &  4   & \cite{Wen_Han_2018} \\
        GALWEIGHT          & 14555   & 1800  &   36 & 3   &  5   & \cite{GALWEIGHT}    \\
        MaDCoWS I          & 27958   & 2433  &   19 & 6   &      & \cite{MaDCoWSI}     \\
        MaDCoWS II         & 6498    & 22970 &  246 & 90  &      & \cite{MaDCoWSII}    \\ 
        LP15               & 25960   &  122  &   13 & 4   &      & \cite{LP15}         \\ 
        REDMAPPER-SDSS-DR8 & 150     & 26111 &  507 & 106 &      & \cite{RSVA-RSDSS}   \\
        REDMAPPER-DES-SVA  & 150     & 1382  &   28 & 1   &      & \cite{RSVA-RSDSS}   \\
        NEURALENS          & 14000   & 1312  &   35 & 8   &      & \cite{NEURALENS}    \\ 
        \bottomrule
    \end{tabular}
    \caption{Public cluster catalogues cross-matched with ComPACT.The catalogues are grouped by their primary observational data, namely SZ (microwave), joint SZ+X-ray, X-ray, and IR/optical surveys. Columns show the survey footprint, the total number of clusters in each catalogue, the number of matches with ComPACT, and the number of clusters for which redshift ($z$) or mass ($M_{500c}$) estimates were added by priority (see Appendix~\ref{app:mass_estimation}).}
    \label{tab:catalogs}
\end{table*}

The ComPACT catalogue spans the ACT survey footprint ($\sim 18,000$ deg$^2$) and contains a total of 2,962 SZ-selected cluster candidates, detected using the ACT+Planck intensity maps with a DL model, of which 1,220 have been validated as galaxy clusters in existing catalogues. These candidates are assigned Priority I–III reliability classes, with Priority I ($N=1,720$) having an expected purity of $\sim84 \%$, as estimated in \cite{Voskresenskaia_2024}. Below, we summarize the multi-wavelength data sets used for cluster confirmation, redshift estimation, and mass calibration:  
\begin{itemize}
    \item \textbf{Optical/infrared imaging (redshift estimates)}: We use deep optical photometry from
    \begin{itemize}
        \item \textbf{the DESI Legacy Imaging Surveys (e.g. DECaLS DR9}, \citealt{DECALS}), which provide optical $g,r,z$ photometry combined with 3.4 and 4.6\,$\mu$m photometry from \textit{WISE} \citep{Wright_2010}. These data are crucial for detecting red-sequence galaxies across most of the ACT cluster search regions.  
        \item \textbf{Wide-field Infrared Survey Explorer (WISE;} \citealt{Wright_2010, Burenin2022}) provides all-sky mid-infrared photometry in $W1$ and $W2$.  
    \end{itemize}

    \item \textbf{Compton y-maps (mass estimates):} For SZ-based mass estimation, we use the publicly available combined ACT+\Planck\ Compton-$y$ map from NASA/LAMBDA\footnote{\url{https://lambda.gsfc.nasa.gov/product/act/act_dr6.02/act_dr6.02_nilc_prod_table.html}}, which covers roughly one-third of the sky with a resolution of $0.5$ arcmin per pixel \citep{y_2024}. We also use the \Planck\ full-mission NILC $y$-map (HEALPix $N_{\rm side}=2048$, 1.72 arcmin per pixel; \citealt{Planck_NILC_y}). For each cluster in the catalogue, we extract its SZ signal from the maps to estimate its mass.

    \item \textbf{ACT+\Planck\ intensity maps (completeness characterization):} To assess a completeness of the ComPACT catalogue, we employ the combined ACT DR5+\Planck\ sky intensity maps at 90, 150, and 220 GHz \citep{Naess_2020} available from LAMBDA,\footnote{\url{https://lambda.gsfc.nasa.gov/product/act/act_dr6.02/act_dr6.02_maps_coadd_get.html}} with a resolution of $0.5$ arcmin per pixel, covering $\sim18,000$ deg$^2$.
\end{itemize}

\thispagestyle{empty}

\section{Redshift measurements}
\label{sec:redshift}
\begin{figure*}
    \centering
    \begin{subfigure}[t]{0.48\textwidth}
        \centering
        \includegraphics[width=\linewidth]{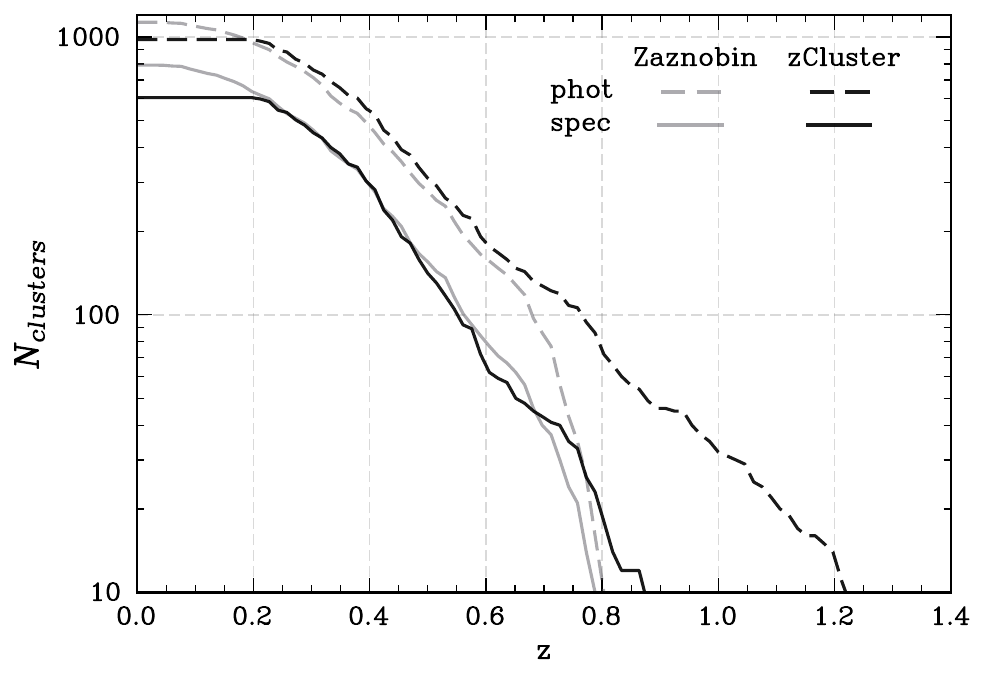}
        \caption{Cumulative redshift distribution of cluster candidates obtained with the \texttt{Zaznobin} method and with \texttt{zCluster}. The \texttt{Zaznobin} method is especially effective for low-redshift systems ($z < 0.2$), whereas \texttt{zCluster} is able to recover clusters out to higher redshifts ($z \gtrsim 0.8$).}
        \label{fig:nz_compare}
    \end{subfigure}
    \hfill
    \begin{subfigure}[t]{0.48\textwidth}
        \centering
        \includegraphics[width=\linewidth]{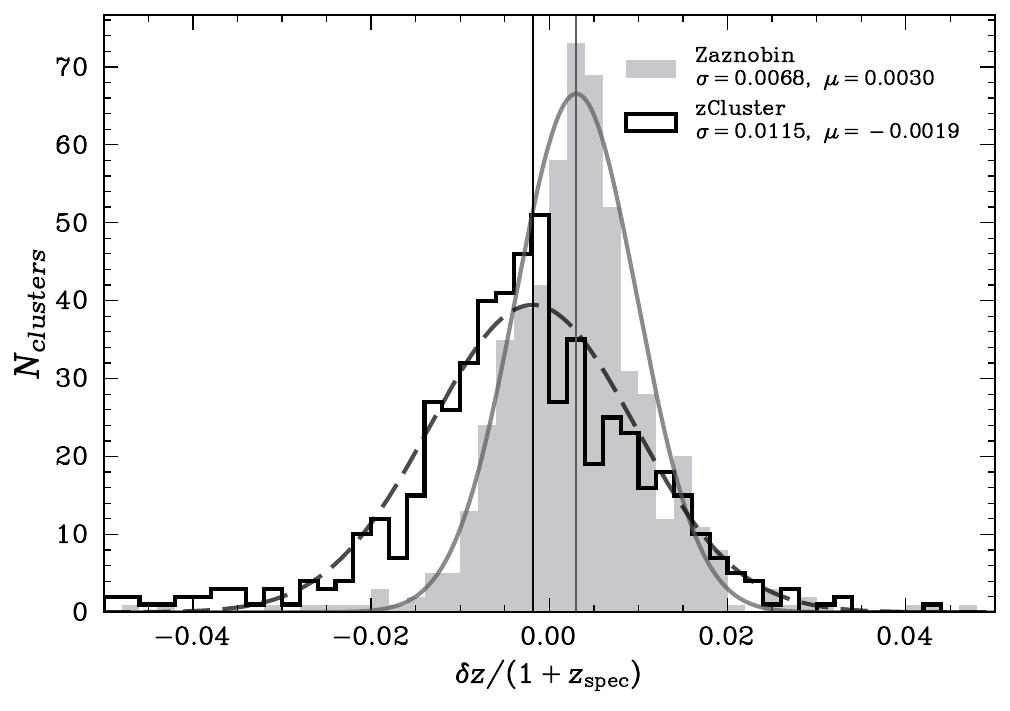}
        \caption{Distribution of the photometric redshift error, $\delta z / (1 + z_{\rm spec})$, for clusters with spectroscopic redshifts.}
        \label{fig:dz_compare}
    \end{subfigure}
    \caption{Comparison of cluster redshift estimates obtained with the \texttt{Zaznobin} method and with \texttt{zCluster}.}
    \label{fig:zcompare}
\end{figure*}
Measuring galaxy cluster redshifts is essential for determining their distances, estimating their masses, and enabling their use in astrophysical analyses. We adopt a hierarchical strategy to obtain redshift information uniformly across all priorities. Redshifts are obtained by cross-matching with external catalogues and by combining the \texttt{zCluster} \citep{Hilton_2018, Hilton_2021} algorithm with the method of \citet{Zaznobin2023} (hereafter \texttt{Zaznobin}; see Section~\ref{subsec:photoz_meth} for details).

\subsection{Literature redshifts}
The ComPACT catalogue is validated through cross-identification with previously known galaxy clusters  from SZ, X-ray, and optical surveys (see Table~\ref{tab:catalogs}). Using a matching radius of 5 arcmin, motivated by the analysis of average radial number density profiles around ComPACT candidates (see Appendix B1 in \citealt{Voskresenskaia_2024}), we identify counterparts for 56\% of the sample (1\,668 clusters). 

Among these associations, 258 correspond to clusters reported in catalogues published after the release of ComPACT, including ACT DR6, ACT-MCMF, SPT-DEEP, MCXC2 (excluding MCXC), and MaDCoWS~II. Redshift information is available in the literature for 1\,656 of the matched systems. When a spectroscopic redshift ($z_{\mathrm{spec}}$) is available, we adopt it as the cluster redshift; otherwise, we use the published photometric redshift ($z_{\mathrm{phot}}$).

\subsection{Photometric redshift estimation for clusters without literature redshifts}
\label{subsec:photoz_meth}

For galaxy clusters without previously reported redshifts, we apply two independent photometric methods: the \texttt{Zaznobin} algorithm and the \texttt{zCluster} method. Both approaches analyse the redshift distribution of galaxies around the cluster centre and identify the most significant overdensity corresponding to the cluster redshift. Our selection scheme is as follows: (i) if a redshift from the \texttt{Zaznobin} method is available, we adopt it; (ii) otherwise, we use the \texttt{zCluster} estimate.

\subsubsection{\texttt{Zaznobin} method}

The algorithm used in this work is a modified version of the method presented by \cite{Zaznobin2023}\footnote{\href{https://github.com/izaznobin/zPhot_Zaznobin}{https://github.com/izaznobin/zPhot\_Zaznobin}}. In the original approach, cluster redshifts were estimated using X-ray data from the \textit{eROSITA} survey, infrared data from \textit{WISE} survey in the W1 band \citep{Wright_2010}, and galaxy photometric redshifts from \cite{2022RAA....22f5001Z}, derived from the DESI Legacy Imaging Surveys DR9 \citep{DECALS}. 

Since X-ray data are not used in the present analysis, the algorithm was adapted to the available data sets. Instead of X-ray information, we used the positions and mass estimates of the SZ sources. When no mass estimate was available, a fiducial mass of $M_{500c} = 10^{14}~M_\odot$ was assumed. This fiducial mass was validated on a spectroscopic subsample: lower thresholds increased sample size but degraded the AUC score (Area Under the Receiver Operating Characteristic Curve) of redshift estimates; the adopted value maximises completeness while maintaining high AUC and minimising redshift scatter.

As in the original work, the method consists of two stages.
\paragraph{Stage 1: Preliminary redshift.}
An optical association of galaxies with SZ sources is performed to obtain a preliminary redshift estimate, $z_{\mathrm{prel}}$. At this stage, galaxies are selected from the catalogue of \cite{2022RAA....22f5001Z} within a predefined spatial region, and the distribution of their infrared luminosities is constructed following the procedure described in \cite{Zaznobin2023}. The main difference with respect to the original method is that the galaxy selection is centred on the SZ source coordinates (RA, Dec) and that the spatial selection criteria are modified. Galaxies are selected within an angular radius of $534$~arcsec from the centre of the SZ source, corresponding to the angular size of $R_{500c}$ for a galaxy cluster of mass $M_{500c}=3\times10^{14}\,M_\odot$ at $z=0.1$. In addition, a cut on the projected physical distance of $814$~kpc is applied, corresponding to $R_{500c}$ for a cluster of the same mass at $z=0.6$.

As in \cite{Zaznobin2023}, the preliminary redshift estimate is defined as the redshift corresponding to the maximum of the infrared luminosity distribution.

\paragraph{Stage 2: Refined redshift.}
Cluster redshift is refined using the same methodology as in \cite{Zaznobin2023}. Galaxies with photometric redshift estimates $\mathrm{photo\_z}$ and corresponding redshift uncertainties $\mathrm{phot\_zerr}$ from the catalogue of \cite{2022RAA....22f5001Z} are selected according to the following criteria:

\begin{enumerate}
    \item projected distance $r < R_{500c}$;
    \item relative photometric redshift uncertainty $\mathrm{phot\_zerr}/(1+\mathrm{photo\_z}) < 2\%$;
    \item $z_{\mathrm{prel}} - 0.06(1+z_{\mathrm{prel}}) < \mathrm{photo\_z} < z_{\mathrm{prel}} + 0.06(1+z_{\mathrm{prel}})$.
\end{enumerate}

For the selected galaxies, a refined cluster redshift is computed as the inverse-variance-weighted mean of their photometric redshifts, with weights  given by the inverse squared photometric redshift uncertainties. The standard deviation of the distribution is computed simultaneously. An iterative $2\sigma$-clipping procedure is then applied: at each iteration, galaxies deviating by more than $2\sigma$ from the mean are removed. The procedure was repeated until no further galaxies are rejected. The final weighted mean redshift is adopted as the refined cluster redshift.

In the original study, the reliability of the cluster identification was quantified using a parameter defined as the product of two factors, $p_1$ and $p_2$ \citep{Zaznobin2023}. In the present work, instead of applying a threshold to their product, independent thresholds are imposed on each factor. The parameters $p_1$ and $p_2$ are computed using a sample of $10^4$ random positions. Since the procedures used to estimate both preliminary and refined redshifts are modified, the reliability parameters are recomputed for the random sample.

Only objects satisfying $p_1 > 0.978$ and $p_2 > 0.8$ are included in the final analysis. These thresholds correspond to a false-positive rate of approximately $5 \%$. After applying this selection, we obtain 47 new photometric redshift estimates.

\subsubsection{\texttt{zCluster} method}

zCluster\footnote{\href{https://github.com/ACTCollaboration/zCluster}{https://github.com/ACTCollaboration/zCluster}}\label{zCluster} is a photometric redshift estimation algorithm for galaxy clusters, originally presented by \cite{Hilton_2018, Hilton_2021}. The method estimates cluster redshifts using broadband photometry, assuming prior knowledge of the cluster position on the sky. In this work, we apply \texttt{zCluster} to photometric data from the DECaLS DR9 survey \citep{DECALS}.

For each galaxy located in the vicinity of the cluster position, an individual photometric redshift probability distribution, $p(z)$, is computed using template fitting. These individual probability distributions are then combined into a single cluster redshift probability distribution through a weighted summation. The weights account for the projected distance of each galaxy from the cluster centre, reflecting the expected radial distribution of galaxies in a cluster, as well as the quality  of the photometric measurements. Consequently, galaxies which are closer to the cluster centre and with more reliable photometric redshift estimates contribute more strongly to the combined probability distribution.

To suppress noise fluctuations and enhance coherent redshift features, the combined probability distribution is smoothed by integrating the individual $p(z)$ distributions over a redshift window, yielding the quantity $n_{\Delta z}(z)$. The photometric redshift of the cluster is then defined as the redshift corresponding to the maximum of this smoothed distribution.

To quantify the optical overdensity associated with the cluster and to assess the reliability of the redshift estimate, \texttt{zCluster} defines a contrast parameter,
\begin{equation}
\delta(z) = \frac{n_{0.5\,\mathrm{Mpc}}(z)}{A\,n_{3\text{--}4\,\mathrm{Mpc}}(z)} - 1 ,
\end{equation}
where $z$ is the estimated cluster redshift, $n_{0.5\,\mathrm{Mpc}}(z)$ is the number of galaxies within a projected radius of 0.5~Mpc from the cluster centre, and $n_{3\text{--}4\,\mathrm{Mpc}}(z)$ is the background galaxy number density estimated within an annulus spanning projected radii of 3--4~Mpc. The factor $A$ accounts for the difference in area between the inner aperture and the background annulus.

The selection of spectral templates used to compute the galaxy photometric redshift probability distributions is calibrated using a randomly selected subsample of 114 ACT DR5 clusters from the ComPACT catalogue. We investigated the impact of different template sets, including the COSMOS galaxy and active galactic nucleus templates \citep{Ilbert_2009, Salvato_2011} and CWW templates \citep{CWW}). Based on this analysis, we adopted a combination of two galaxy templates (Sb\_A\_0, Sb\_template\_norm) for the \texttt{zCluster} $\delta$ measurement and redshift determination at $z>0.2$, as this choice yields the best performance in terms of photo-z dispersion and the fraction of catastrophic outliers  for the ComPACT clusters with available spectroscopic redshifts.

Following the approach adopted for the main redshift estimation algorithm, the reliability of the \texttt{zCluster} redshift measurements was assessed using the distribution of $\delta$ measured at random positions (see Appendix~\ref{zCluster_thr}). Based on this analysis, we adopted a threshold corresponding to a false-positive rate of 5\%, which translates into a requirement of $\delta \geq 3$. Only objects satisfying this criterion were included in the final sample. After applying this selection, we obtained 69 new photometric redshift measurements.

\subsection{Comparison of photometric methods}

Figure~\ref{fig:zcompare} compares the cumulative redshift distributions and photometric redshift performance for clusters with available spectroscopy. 

The \texttt{Zaznobin} method achieves $\mu = 0.0030$, $\sigma = 0.0068$, and an outlier fraction $|\delta z|/(1+z) > 20\%$ of 0.011. The \texttt{zCluster} method yields $\mu = -0.0019$, $\sigma = 0.0115$, and an outlier fraction of 0.007. \texttt{Zaznobin} provides higher precision at low redshift ($z<0.8$), while \texttt{zCluster} extends to higher redshifts ($z\gtrsim0.8$) with a slightly lower catastrophic outlier fraction.

When compared directly on the spectroscopic subsample, the two methods show excellent agreement: the normalized median offset is $\mathrm{median}[\Delta z/(1+z)] = -0.0051$, with a robust scatter of $\sigma_{\mathrm{NMAD}} = 0.0111$. This consistency between independent methodologies supports the robustness of our cluster redshift estimates.

\subsection{Final redshift sample and completeness}

The resulting sample of 1771 galaxy clusters with measured redshifts (spectroscopic or photometric) has a mean redshift of $\langle z \rangle = 0.43$, comparable to other SZ-selected samples, and covers the range $0.007 < z < 1.795$. For the ComPACT catalogue, the total number of clusters with spectroscopic redshifts is 1,027 clusters ($\sim 34$ \%), while the photometric redshifts are available for an additional 628 (21.2 \%) clusters. Our photometric analysis provides 116 new redshift measurements. 
The redshift completeness depends on the catalogue priority: the Priority~I subsample reaches 76\% completeness, compared to 44.6 \% for Priority II and 33.5 \% for Priority III. The overall redshift completeness of the full ComPACT catalogue is 60 \%.

\section{Mass estimation}
\label{subsection:mass}
Cluster masses are derived using SZ measurements from the ACT+\Planck\ and \Planck\ Compton-$y$ maps, complemented by values available in the literature.

To derive mass estimates for clusters, we use the following approaches:
\begin{enumerate}
    \item We cross-match our candidates with previously published catalogues (see Table~\ref{tab:catalogs}) and extract available SZ, X-ray, and optically-derived mass estimates. As these  estimates are affected by different systematics and calibration biases, we include the \textit{mSource} column to specify the origin of each mass estimate. Selecting cluster candidates based on \textit{mSource} allows one to construct subsamples with masses defined consistently across the sample.
    \item For unmatched objects, we detect these cluster candidates in the ACT+\Planck\ y-map by computing the signal-to-noise ratio. Prior to signal extraction, the ACT+\Planck\ y-map, originally in plate-carée projection with 0.5 arcmin per pixel, is reprojected to the HEALPix format with \texttt{Nside=8192} to preserve the native pixel scale. To calculate the signal, we define a circular region with a radius of $r = 1.6$ arcmin and compute the maximum tSZ signal within this aperture. To estimate the noise, we follow \cite{SNR_ACT2024} and  use a background annulus around each candidate. The annulus has an inner radius of $R_1 = 30 \times r$, measured from the candidate centre, and an outer radius of $R_2 = R_1 + 30'$. Within this annulus, we measure the maximum tSZ signal in 1,000 randomly placed circular apertures (each 1.6 arcmin in radius) and compute the mean ($y_{bkg}$) and standard deviation ($\sigma_{bkg}$) of these values. The SNR is given by:
    \begin{equation}
    \label{eq:snr}
    \mathrm{SNR} = \frac{y_{max} - y_{bkg}}{\sigma_{bkg}}
    \end{equation}
    
    Clusters are selected with SNR\footnote{As a consistency check, we examined the correlation between the derived SNR and the availability of mass estimates in the literature using the Spearman rank correlation test. We find a statistically significant positive correlation between SNR and the availability of literature mass estimates for both the ACT and Planck subsamples} $> 2$ corresponding to 5\% of false detections (for details, see~App.~\ref{sec:yact_threshold}). To estimate the mass, we compute the "cylindrical" integrated $Y^{cyl}$ parameter within an aperture of $R = 10'$ to ensure consistency with \Planck\, whose map resolution (FWHM) is  approximately 10~arcmin.  
    The measured $Y^{\mathrm{cyl}}$ is scaled by a factor $a$ to match \Planck\ measurements. We derive $a = 0.8$ by matching our $Y$-values to \Planck\ $Y$-measurements for a calibration set of PSZ2 clusters processed with identical apertures.
    $Y^{\mathrm{cyl}}$  is then converted to the "spherical" $Y_{500}$ following the procedure described by \cite{Melin_2011}, which assumes a gNFW pressure profile, and finally, the mass $M_{500c}$ is derived using the scaling relation from \cite{Planck_2014} (for details, see App.~\ref{app:mass_estimation}).

    \item For regions not covered by the ACT+\Planck\ y-map, we use the \Planck\ y-map. Cluster detection is performed using the same background-annulus methodology as for the ACT+\Planck\ y-map. The aperture radius of $R = 1.6'$ was selected by maximising the recovery fraction of PSZ2 clusters (yielding a completeness of 89\%) at a fixed false detection rate of 5\%, corresponding to a detection threshold of $\mathrm{SNR} > 1.7$.
    This angular scale also closely matches the native HEALPix pixel size of the \Planck\ NILC map ($\sim$1.72' for Nside=2048). The mass is then estimated from the measured $Y$-values using the same scaling relation as above.
\end{enumerate}

We obtain mass measurements for 56\% of the full catalogue and for 72.9\% of the priority I subsample. The average mass of the ComPACT sample with available mass estimates is $\langle M_{500c} \rangle = 4 \times 10^{14}\, M_\odot$, with values ranging from $0.25 \times 10^{14}$ to $13.1 \times 10^{14}\, M_\odot$. In total, we provide 158 new mass measurements: 114 derived from ACT+\Planck\ and 44 from \Planck\ y-maps.

\section{Results}
\label{sec:results}
\thispagestyle{empty}
\subsection{Completeness}
\label{subsec:completeness}
\begin{figure*}
    \centering
    \begin{subfigure}[t]{0.48\textwidth}
        \includegraphics[width=\linewidth]{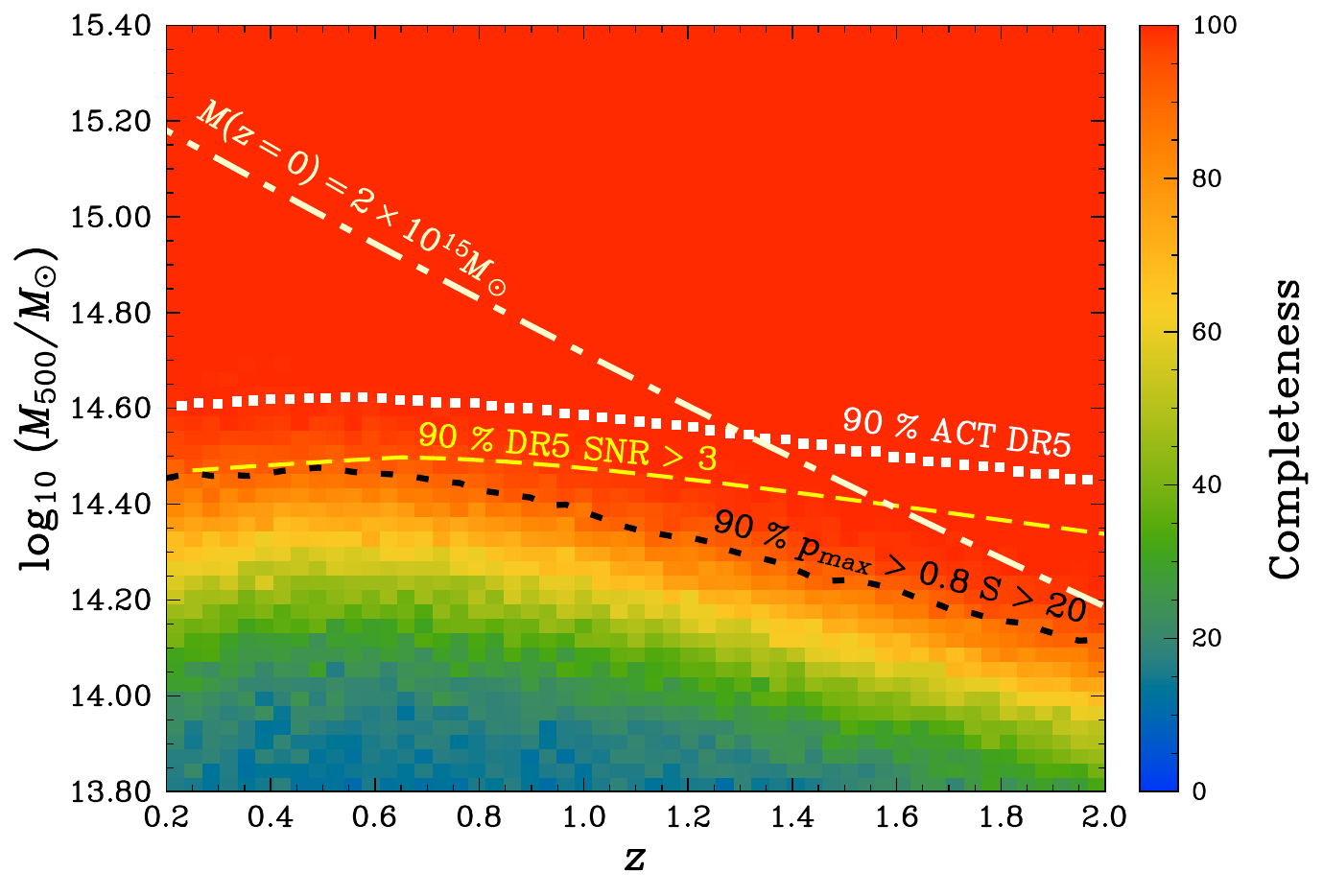}
        \caption{Within the ACT DR5 footprint ($13,211$ deg$^2$). The black dashed contour shows the 90\% completeness of the catalogue, while the solid yellow line marks the same level for the ACT DR5 $SNR>3$ catalogue at matched purity. The ComPACT achieves higher completeness than ACT DR$5_{SNR>3}$ at fixed purity, especially for massive clusters at high redshift, while both catalogues perform comparably at low $z$.}
        \label{fig:completeness_act}
    \end{subfigure}
    \hfill
    \begin{subfigure}[t]{0.48\textwidth}
        \centering
        \includegraphics[width=\linewidth]{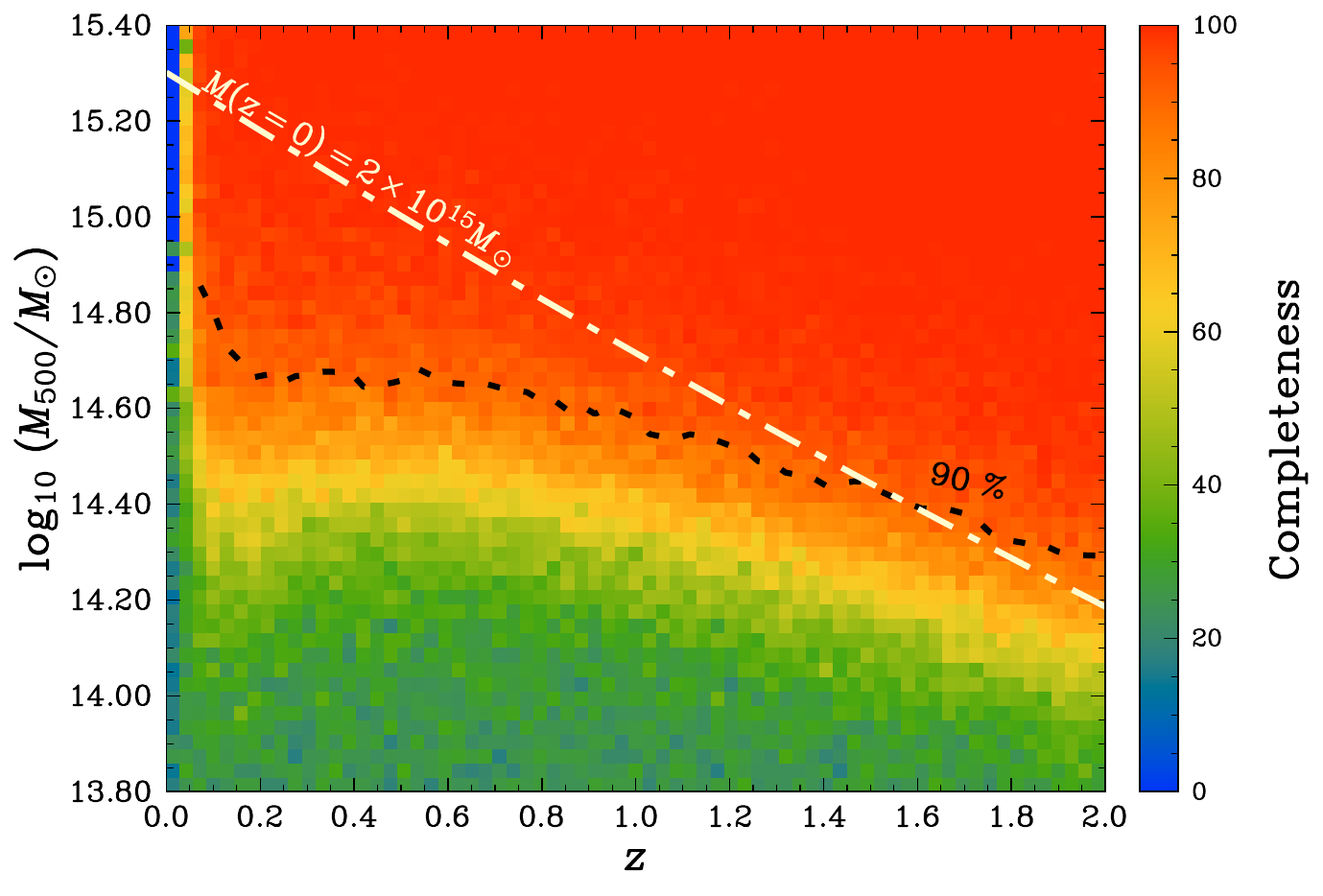}
        \caption{Outside the ACT DR5 footprint ($\sim3000$ deg$^2$). Catalogue retains high completeness for massive clusters across the redshift range, demonstrating its potential to identify new systems beyond existing survey coverage.}
        \label{fig:completeness_notact}
    \end{subfigure}
    \caption{Estimated completeness of the DL based cluster catalogue {(without \Planck\ SZcat selection)} in the $(M_{500c}, z)$ plane. The black dashed line corresponds to detection thresholds of $p_{\max}>0.8$ and $S>20$. The white dash-dotted line represents the average mass evolution of a $2 \times 10^{15} M_\odot$ cluster over redshift \citep[from ][]{Fakhouri_2010}.}
    \label{fig:completeness_combined}
\end{figure*}

Completeness of SZ-catalogues is typically estimated by injecting the UPP model clusters into maps and measuring the recovery rates \citep{2013ApJS..209...17H, 2016A&A...594A..27P}, as implemented, for example, in the ACT DR5 analyses \citep{Hilton_2018}. We assess the ComPACT catalogue completeness in a similar way by injecting synthetic clusters are injected into the real ACT+\Planck\ maps and passing them through the deep-learning detection pipeline (see Appendix~\ref{sec:simul} for details). Figure~\ref{fig:completeness_act} shows the completeness in the $(M_{500c}, z)$ plane for detections with $p_{\max}>0.8$ and $S>20$ (see fig. 5 in \cite{Voskresenskaia_2024}) without selection in SZcat directions. Here, $p_{max}$ denotes the maximum classification probability assigned by the neural network to a detected object, while $S$ is the area of the detection in pixel units; both quantities characterize objects after a classification. These selection thresholds are applied after the classification stage and the analysis is restricted to the ACT DR5 footprint ($13,211$ deg$^2$). 

To enable a direct comparison with the cumulative completeness estimate reported in \citet{Voskresenskaia_2024}, we also compute the integrated completeness from our simulations for clusters with $M_{500c} > 10^{14} M_\odot$ across the full redshift range. We obtain a cumulative completeness of 77 \%. This value differs from the $\approx 70\%$ reported in \citet{Voskresenskaia_2024} because our simulation-based estimate does not include the selection by SZcat directions applied to the ComPACT catalogue; consequently, the $77\%$ value should be interpreted as an upper limit on the ComPACT completeness.

For a fair comparison with ACT DR5, we matched the catalogues at equal purity. Using the \textsc{Nemo}\footnote{\url{https://github.com/simonsobs/nemo/}} package, we obtained the ACT DR5 $SNR>3$ sample (yellow 90\% contour in Fig.~\ref{fig:completeness_act}), corresponding to a purity of $\sim20\%$, consistent with the full catalogue prior to SZcat-based selections. We note here that the mass scales of the ComPACT and ACT DR5 catalogues are aligned by construction, since both analyses use the same input $  M_{500c}$ distribution of synthetic clusters. At low redshift, both catalogues exhibit similar completeness, but the DL approach recovers more massive systems at higher redshifts. Overall, the ComPACT catalogue achieves higher completeness than ACT DR5 at fixed purity. This behaviour indicates that the deep-learning approach has improved sensitivity to lower signal-to-noise systems compared to traditional matched-filter methods, particularly at high redshift.

The dash-dotted curve shown in Fig.~\ref{fig:completeness_act} represents the median mass accretion history of dark matter haloes from \citet{Fakhouri_2010}. It is included as a reference to illustrate the expected evolutionary relation between cluster mass and redshift and to guide the interpretation of the completeness limits in the ($M_{500c}, z$) plane and to distinguish the population of the most massive clusters.

The same analysis is repeated for the ComPACT survey region, excluding ACT DR5 field to avoid overlap ($\sim3000$ deg$^2$). Outside the footprint of ACT DR5 catalogue (see Fig.~\ref{fig:completeness_notact}), our catalogue demonstrates higher completeness for massive galaxy clusters across the entire redshift range. This indicates that in these regions, the most massive cluster candidates may be discovered, and this prediction is borne out by our optical/NIR follow-up observations (Sec.~\ref{subsec:mz-describe}).

\subsection{Confirmation of clusters}
\label{purity}
\begin{table*}[ht!]
    \centering
    \begin{tabular}{l|c|c|c|c|c|c|c|c}
        \hline\hline
        & \multicolumn{4}{c|}{ACT DR5 footprint} & \multicolumn{4}{c}{DESI  LIS footprint} \\
        & \multicolumn{2}{c|}{Inside} & \multicolumn{2}{c|}{Outside} & \multicolumn{2}{c|}{Inside} & \multicolumn{2}{c}{Outside} \\
        \cline{2-9}
        & Number        & Percentage  & Number & Percentage  & Number & Percentage & Number & Percentage    \\
        &        &  (Purity) &  &  (Purity) &  &  (Purity) &  &  (Purity)   \\
        \hline
        Total              & 2587 &              & 375 &              & 2448 &              & 514 & \\ 
        Confirmed          & 1640 & 63.4 ~(63.2) & 144 & 38.4 ~(38.2) & 1531 & 62.5 ~(62.3) & 253 & 49.2 (49.2)\\
        Redshift estimated & 1634 & 63.2 ~(63)   & 137 & 36.5 ~(36.3) & 1525 & 62.3 ~(62.1) & 246 & 47.9 (47.9)\\
        Mass estimated     & 1527 & 59.4 ~(58.8) & 125 & 32.5 ~(32.5) & 1416 & 57.7 ~(57) & 246 & 47.9 (47.9)\\
        New (with z\&$M_{500c}$)    & 58   & 2.2  ~(2.1)  & 10   & 2.7  ~(2.5)  & 72   & 2.9 ~(2.6)   & 0    &\\
        \hline
        \hline
        \end{tabular}
    \caption{Number and percentage of candidates that are classified in the four different classes (confirmed, with redshift estimates, with mass estimates and new with redshift and mass estimates) in different regions: inside and outside the ACT DR5 and DESI LIS footprint.}
    \label{tab:data_distribution_ACT}
\end{table*}
\begin{table*}[ht!]
\centering
        \begin{tabular}{l|c|c|c|c|c}
        & \multicolumn{4}{c|}{Catalogue footprint} & \\
        & \multicolumn{2}{c|}{}&\multicolumn{2}{c|}{Priority 1}\\
        \cline{2-5}
                                     & Number & Percentage (Purity) & Number & Percentage (Purity)\\
        \hline
        Total                        & 2962 &               &  1720 &\\
        Confirmed                    & 1784 &  60.2 ~(60)   &  1308 & 76 ~(75.9)\\
        - Cross-matched              & 1668 &               &       &           \\
        \hline
        Redshift estimated           & 1771 &  59.8 ~(59.6) & 1300 & 75.6 ~(75.5)\\
        \quad  - Cross-matched       & 1655 &  55.9 & &\\
        \quad  ~   $+$ spec          & 1027 &  31 & &\\
        \quad  ~   $+$ phot          & 628  &  22 & &\\
        \quad  - \texttt{Zaznobin}   & 1130 &  38.1  &&\\
        \quad  - \texttt{zCluster}   & 979  &  33.1  &&\\
        \hline
        Mass estimated               & 1663 & 56 ~(55.9)  & 1257 & 73.7 ~(73) \\
        \quad  - Cross-matched       & 1504 & 50.7  &       &           \\
        \quad  - yACT+\Planck\       & 1441 & 48.6  &       &           \\
        \quad  - yPlanck             & 1378 & 46.5  &       &           \\
        New                          & 116  &  3.9  &  18   & 1.1  ~(1)\\
        \quad  - \texttt{zCluster}   & 69   &  2.3  &       &           \\
        \quad  - \texttt{Zaznobin}   & 47   &  1.6  &       &           \\
        \quad  - yACT+\Planck\       & 114  &  3.8    &       &           \\
        \quad  - yPlanck             & 44   &  1.4  &       &           \\
        \quad  - new z\&$M_{500c}$   & 72   &  2.3 ~(2.2) &  26   &  1.5  ~(1.4)\\
        
    \end{tabular}
    \caption{Number and percentage of candidates that are classified in the three different classes in full catalogue and in priority I. }
    \label{tab:data_distribution}
\end{table*}

Cluster candidates are considered as confirmed clusters if they satisfy at least one of the following criteria:
\begin{enumerate}
    \item They have a cross-match with galaxy cluster catalogues listed in  Table~\ref{tab:catalogs}. In total, 1,668 clusters have such counterparts.
    \item They are  identified by the \texttt{zCluster} algorithm with an optical contrast of  $\delta > 3$ (69 objects), or their \texttt{Zaznobin} score $p_1 > 0.978$ and $p_2 > 0.8$ corresponds to 5 \% false-positive rate (47 objects) (see Section~\ref{sec:redshift}); 
\end{enumerate}

Table \ref{tab:data_distribution_ACT} summarises the number and fraction of cluster candidates — together with the corresponding purity — classified as confirmed, with redshift estimates, with mass estimates, or newly identified with redshift and mass estimates, both inside and outside the ACT DR5 and DESI Legacy Imaging Surveys (LIS; DeCALS DR9) footprints. The fraction of confirmed clusters is significantly higher inside the ACT DR5 footprint (63.2\%) than outside (38.2\%). A similar trend is observed for the DESI LIS footprint, where 62.5\% of candidates inside the footprint are confirmed, compared to 49.2\% outside. This decrease of approximately 10\% in the confirmation rate outside the survey footprints is primarily driven by fewer photometric data from DESI LIS. As a consequence, we expect that future redshift follow-up efforts may increase the confirmed fraction outside the ACT DR5 footprint by up to $\sim$10 \%.

Table \ref{tab:data_distribution} provides a summary of the number and fraction of sources in the full catalogue and in the Priority I subsample. As expected, the Priority I subset exhibits the highest reliability: 76 \% of its candidates are confirmed, compared to 60.2\% in the full sample. Redshift estimates are obtained for 59.8\% of all candidates. The \texttt{Zaznobin} method yields larger number of redshift determinations, exceeding those derived with \texttt{zCluster}. Mass estimates are available for 56\% of the full catalogue. Using the ACT+\Planck\ Compton-$y$ maps, masses can be derived for 48.6\% of the candidates, while the \Planck-only $y$ maps allow mass measurements for 46.5 \%. We also identify 116 previously unreported systems, among which 72 have both redshift and mass estimates. The Priority I subsample contributes 26 such new clusters.

\begin{figure*}
    \centering
    \includegraphics[width=0.9\linewidth]{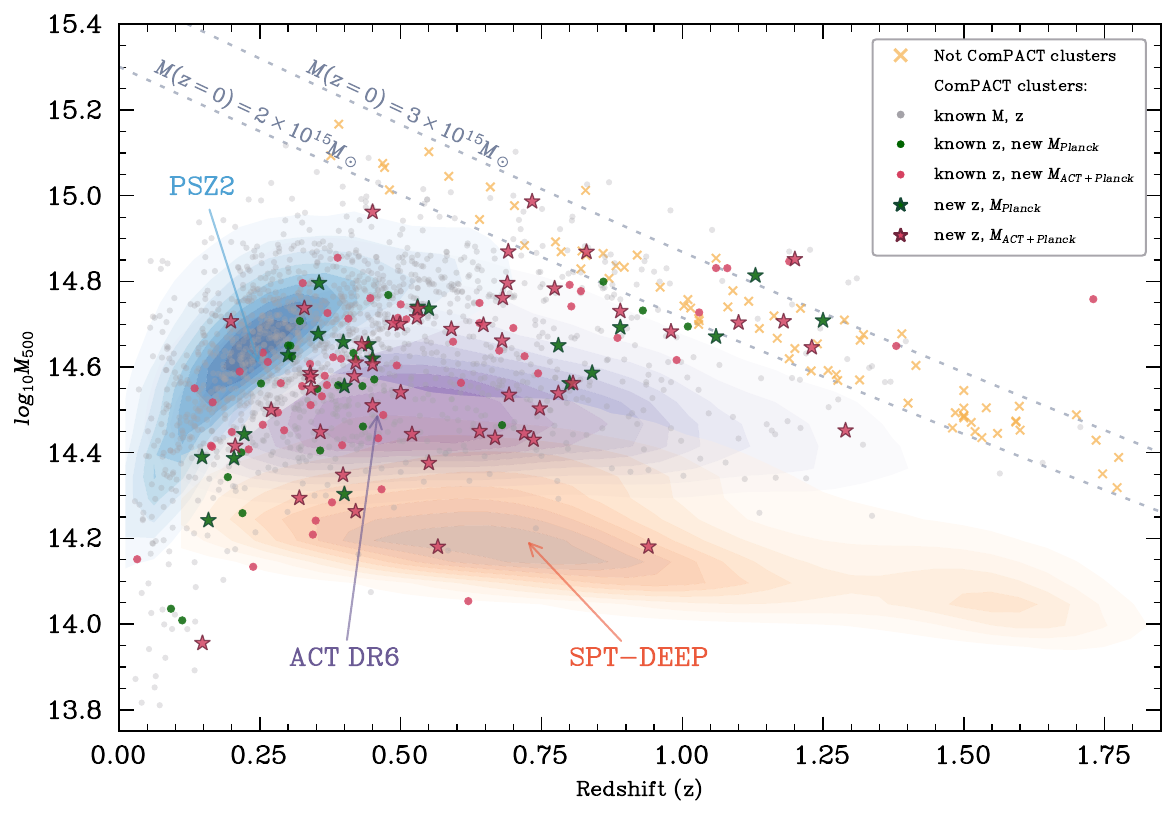}
    \caption{Comparison of the ComPACT cluster sample in the mass–redshift plane. The distribution of ComPACT clusters is shown alongside those from ACT DR6 (purple contours), PSZ2 (blue contours), and SPT-DEEP (orange contours). The dashed lines trace the mass evolution of $2$ and $3\times10^{15}~M_\odot$ clusters at $z=0$ for a fixed cosmology \citep{Fakhouri_2010}. Circles indicate clusters with redshifts from the literature (Table~\ref{tab:catalogs}), while stars denote clusters with newly determined redshifts. Colours distinguish cluster categories: grey points mark ComPACT clusters with literature mass measurements, green points indicate new mass estimates from \Planck\ y-maps, and magenta points show new mass estimates from ACT+\Planck\ y-maps. Clusters with newly determined redshifts and masses are highlighted in green. Orange points above the $2\times10^{15} M_\odot$ line represent clusters from the catalogues listed in Table~\ref{tab:catalogs} that are not included in ComPACT.}
    \label{fig:m-z}
\end{figure*}

\subsection{Mass-redshift behaviour}
\label{subsec:mz-describe}
Figure~\ref{fig:m-z} shows the mass–redshift distribution for ComPACT clusters. For comparison with existing surveys, we overlay contours from the ACT DR6 catalogue (purple), the PSZ2 catalogue (blue) and SPT-DEEP (orange). Known clusters from external catalogues included in ComPACT (see Table~\ref{tab:catalogs}) are shown as dots, while stars denote objects with newly determined redshifts. Colours indicate the source of the mass estimate: grey for masses from literature, green for $M_{500c}$ from the \Planck\ y-map, and magenta for masses from the ACT+\Planck\ y-map.

As evident from the figure, a significant fraction of new ComPACT clusters lies below the PSZ2 distribution and above the ACT DR5 sample, especially at high redshifts ($z > 0.8$). In this region, many clusters have newly obtained estimates for both redshift and mass.

\subsection*{Performance of massive clusters}
\label{susec: notable clusters}
To assess the contribution of ComPACT at the high-mass end, we apply a mass threshold of $2 \times 10^{15}\, M_\odot$ to all clusters at $z > 0.8$ in known X-ray and SZ catalogues. In Figure~\ref{fig:m-z}, orange dots mark all such massive clusters from catalogues listed in Table~\ref{tab:catalogs}. Under this cut, ComPACT recovers approximately 30\% of these systems, with about 10\% being previously unidentified clusters.  

We highlight several of the most massive ComPACT systems, identified through \texttt{zCluster} detections, SZ measurements, catalogue cross-matching (e.g., \texttt{SIMBAD}), and visual inspection of optical and IR data. The catalogue includes the well-known Bullet Cluster (ComPACT\_G266.020$-$21.244) and El Gordo (ComPACT\_G297.972$-$67.759), along with newly identified high-mass systems. 

\begin{figure*}
    \centering
    \includegraphics[width=1\linewidth]{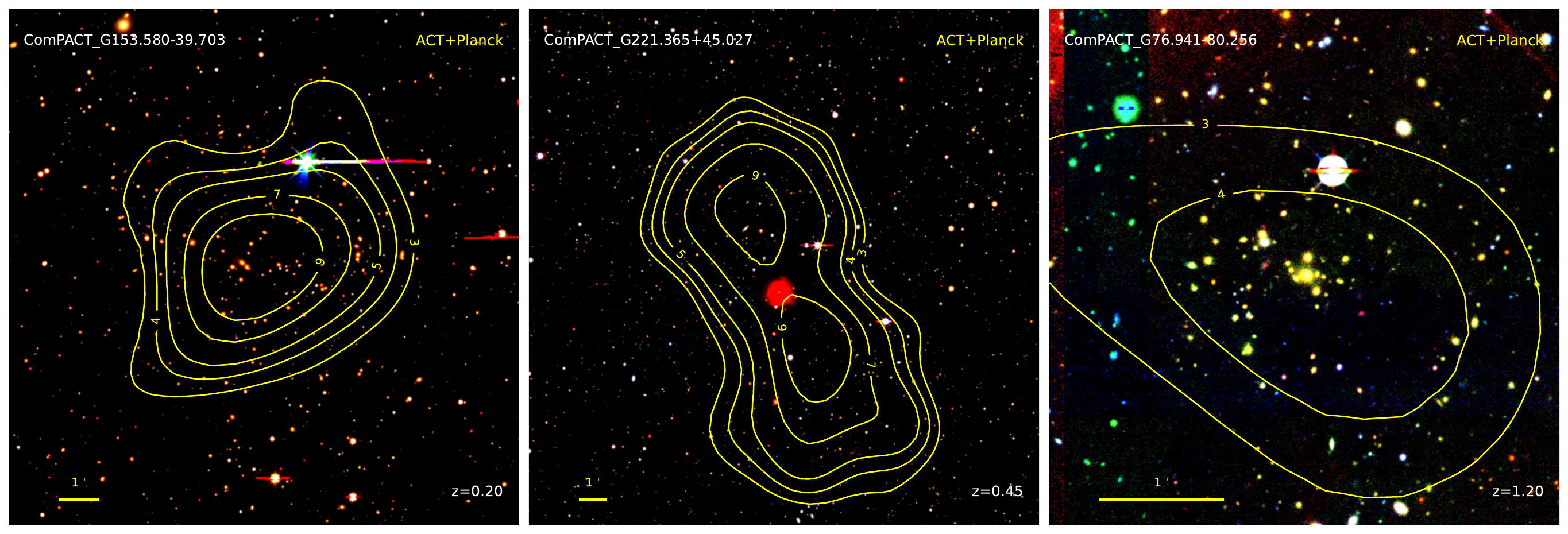}
    \caption{Most massive clusters that are detected by \texttt{zCluster} or \texttt{Zaznobin} algorithms on z = 0.2, 0.69 and 1.2. Each panel shows an optical RGB image from the DESI Legacy Imaging Surveys with overlaid SZ–detection contours. For y-maps, contour levels correspond to $\mu + \sigma \times \{3, 4 \dots\}$, where $\mu$ and $\sigma$ are estimated locally within an annulus around each cluster, as described in Sec.~\ref{subsection:mass}. Contour labels indicate the corresponding significance level in units of $\sigma$.}
    \label{fig:Optical_SZ_image}
\end{figure*}

In Table~\ref{tab:mz_highz}, we present 14 massive cluster candidates that are detected by \texttt{zCluster} or \texttt{Zaznobin} algorithm, including 11 newly identified objects as well as previously known clusters (see the notes column). Optical images of the clusters are shown in Figure~\ref{fig:Optical_SZ_image}. As noted by \cite{2021AstL...47..443B}, not all massive \Planck\ clusters exceed the detection threshold of the PSZ2 catalogue. This can be explained by selection effects in SZ surveys, such as template mismatch and reduced sensitivity to systems with large angular extent or lower SZ signal (e.g., \cite{Lin2021}). Several such clusters are included in Table~\ref{tab:mz_highz}.

\begin{table*}
\centering
\begin{tabular}{lcccccccl}
\hline
Name & Priority & $z$ & $\delta$ & $p_1$ & z Source & $M_{500c}\,$        & Mass  & Notes \\
     &          &     &          &       &          & $[10^{14} M_\odot]$ &  Source    &        \\
\hline
\hline
G75.828$-$51.800  & 1 & 1.73 & ---  & ---  & MaDCoWS II         & 5.7 & ACT+\Planck\     & MOO2 J23036+00542 \\
G183.567$-$76.541 & 2 & 1.29 & 3.61 & ---  & \texttt{zCluster}  & 4.4 & ACT+\Planck\     &  \\
G208.227$+$43.544 & 2 & 1.25 & 11.3 & ---  & \texttt{zCluster}  & 5.1 & \Planck\     &  \\
G253.752$+$68.133 & 3 & 1.23 & 3.57 & ---  & \texttt{zCluster}  & 4.4 & ACT+\Planck\ & \\
G76.941$-$80.256  & 1 & 1.20 & 3.28 & ---  & \texttt{zCluster}  & 7.1 & ACT+\Planck\ & \\
G122.344$-$58.320 & 1 & 1.19 & 4.07 & ---  & MaDCoWS II         & 7.0 & ACT+\Planck\ & MOO2 J00501+04337\\
G241.779$-$81.903 & 3 & 1.18 & 5.16 & ---  & \texttt{zCluster}  & 5.0 & ACT+\Planck\ & \\
G16.016$+$74.451  & 1 & 1.13 & 8.68 & ---  & \texttt{zCluster}  & 6.5 & \Planck\     & \cite{SRGe_J135528.4+214254}\\

G347.312$-$69.060 & 2 & 0.83 & 3.29 & ---  & \texttt{zCluster}  & 7.4 & ACT+\Planck\ &  \\
G135.851$-$52.303 & 1 & 0.73 & ---  & 0.99 & \texttt{Zaznobin}  & 9.7 & ACT+\Planck\ &  \\
G269.871$-$47.295 & 1 & 0.69 & ---  & 0.99 & \texttt{Zaznobin}  & 7.4 & ACT+\Planck\     &  \\

G221.365+45.027   & 2 & 0.45 & 4.84 & ---  & \texttt{zCluster}  & 9.1 & ACT+\Planck\ &  JCMTSE J094747.2+132044 \\
   &  &  &  &   &   &  &  & \citep{SCUBA}\\

G74.431$-$29.961  & 3 & 0.38 & ---  & 0.987  & REDMAPPER        & 7.2 & ACT+\Planck\     & \object{[RRB2014]} \\

G153.580-39.703   & 1 & 0.19 & ---  & 0.99 & \texttt{Zaznobin}  & 5   & ACT+\Planck\ &  \object{PSZRX\%20G153.58-39.69}\\

\end{tabular}
\caption{Sample of massive galaxy clusters detected by the \texttt{zCluster} or \texttt{Zaznobin} algorithms. The table is divided into two sections. The top section lists high-redshift ($z \gtrsim 1.1$) clusters with a present-day equivalent mass $M(z=0) > 2 \times 10^{14} , M_\odot$. The bottom section contains the most massive clusters identified at lower redshifts, $z < 0.8$. The sources for the redshift and mass estimates are provided in the corresponding columns. Associated identifiers and literature references are given in the Notes column.}
\label{tab:mz_highz}
\end{table*}

\section{Conclusions}
\label{sec:conclusions}
We present new mass and redshift measurements for the DL based ComPACT galaxy cluster catalogue. This catalogue expands the SZ cluster population by identifying previously unknown systems. The main results are summarized below:
\begin{enumerate}
    \item By simulating clusters in ACT+\Planck\ maps, we assessed the catalogue completeness across the "mass–redshift" plane. At fixed purity, the catalogue achieves higher completeness than ACT DR$5_{SNR>3}$ catalogue, particularly improving the recovery of massive high-redshift clusters;
    
    \item The catalogue contains 2,962 SZ sources. Cross-matching with existing SZ, X-ray, and optical/IR cluster catalogues yields 1,668 associations. Using DESI LIS surveys photometry, we optically confirm 116 new clusters;

    \item Redshifts are obtained for 1,771 clusters (60 \% of the sample), including 116 new measurements. Masses are estimated for 1,659 objects (56 \%), with 158 new determinations. The median mass is $M_{500c} \sim 4 \times 10^{14} M_\odot$, median $z \sim$0.43;

    \item We found 5 massive ($M_{500c} \gtrsim 6 \times 10^{14} M_\odot$) and distant ($z$ > 0.7) clusters.
\end{enumerate}

The ComPACT catalogue, based on a neural network technique, provides a valuable resource for future studies of galaxy clusters in combination with AI. 

\begin{acknowledgements}
This work was supported by the Russian Science Foundation (Project № 25-22-00470).

We acknowledge the publicly available software packages that were used throughout this work: NumPy \citep{Oliphant:06, Walt:11, Harris:20}, pandas \citep{pandas:2023, mckinney-proc-scipy-2010}, Matplotlib \citep{Hunter:2007}, Astropy \citep{astropy13, astropy_2018, astropy_2022}, pixell \href{https://github.com/simonsobs/pixell}{https://github.com/simonsobs/pixell}, Core Cosmology Library \citep{Chisari_2019}, HEALPix package \citep{Gorski_2005ApJ...622..759G}.

We acknowledge the use of the Legacy Archive for Microwave Background Data Analysis (LAMBDA), part of the High Energy Astrophysics Science Archive Center (HEASARC). HEASARC/LAMBDA is a service of the Astrophysics Science Division at the NASA Goddard Space Flight Center. This research is based on observations obtained with \Planck\ (\href{http://www.esa.int/Planck}{http://www.esa.int/Planck}), an ESA science mission with instruments and contributions directly funded by ESA Member States, NASA, and Canada. This research has made use of the SIMBAD database, operated at CDS, Strasbourg, France.
\end{acknowledgements}



\bibliographystyle{aa}
\bibliography{clusters} 



\begin{appendix}

\section{Description of the new ComPACT catalogue columns}

Table~\ref{tab:cluster_data} describes the contents of the ComPACT catalogue which is publicly available at \href{https://github.com/astromining/ComPACT}{https://github.com/astromining/ComPACT}. The data will be also  made available through at the VizieR sevice.

\begin{table}[ht!]
    \centering
    \begin{tabular}{l|l}
        \hline \hline
        Column & Description \\ 
        \hline
        name & Cluster name in format: \\
             & ComPACT\_JHHMM$\pm$DDMM \\ 
        RA & Right Ascension in decimal degrees (J2000)  \\ 
               & of the SZ detection \\ 
        DEC & Declination in decimal degrees (J2000)\\
               & of the SZ detection \\ 
        pmax & SZ signal probability \\ 
        S & Object area on the SZ signal \\
        &  segmentation map in pixels \\
        Priority & Object priority categorized by SZ signal \\ 
        & and probability \\
        z & Cluster redshift \\
        zType & Redshift type (spec = spectroscopic, \\
        & phot = photometric) \\ 
        zSource & Source of  the cluster redshift (see Table~\ref{tab:catalogs}) \\ 
        zCluster\_delta & \texttt{zCluster} density contrast statistic \\
                        & uncertainty column: zCluster\_err \\
        zZazn\_sig1 & \texttt{Zaznobin} first significance value\\
        zZazn\_sig2 & \texttt{Zaznobin} second significance value\\
                   & uncertainty column: zZazn\_err \\
        M500 & $M_{500c}$ in units of 10$^{14}$ M$_\odot$ \\
             & uncertainty column: E\_M500 (upper bound), \\
             & ~~~~~~~~~~~~~~~~~~~~~~~~~~~~~~~~~e\_M500 (lower bound)\\ 
        mSource & Source of mass (see Table~\ref{tab:catalogs}) \\ 
        Mact & M estimated from Y-M relation from  \\
             & y-ACT+Plank maps (see sec.~\ref{subsection:mass}) \\
             & uncertainty columns: e\_Mact, E\_Mact \\ 
        Mplanck & M estimated from Y-M relation from \\
             & y-Planck maps (see sec.~\ref{subsection:mass}) \\
             & uncertainty columns: e\_Mplanck, E\_Mplanck \\
        \hline
    \end{tabular}
    \caption{Description of the ComPACT catalogue columns.}
    \label{tab:cluster_data}
\end{table}

\section{Detection thresholds}
\label{app:thresholds}
\begin{figure*}
    \centering
    \begin{subfigure}[t]{0.32\textwidth}
        \centering
        \includegraphics[width=\linewidth]{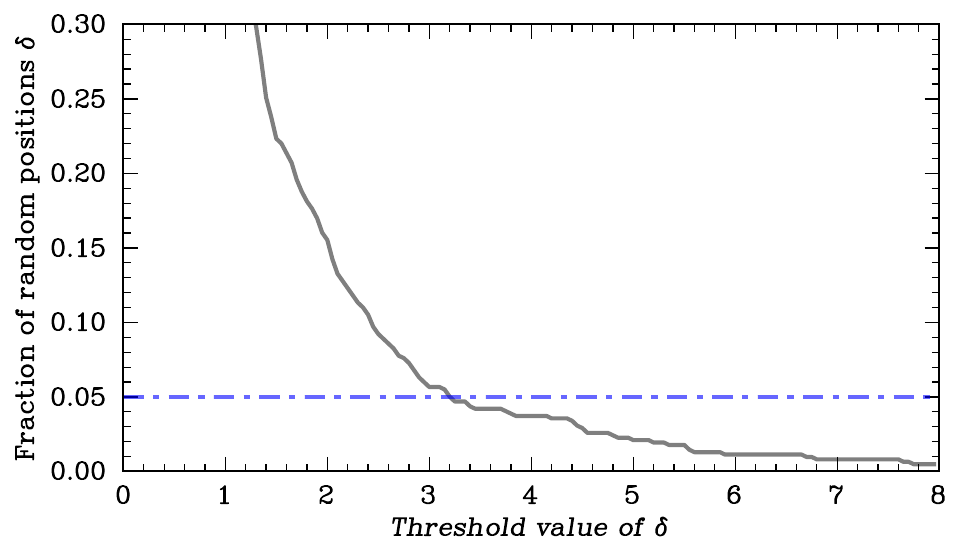}
        \caption{Distribution of the \texttt{zCluster} density contrast statistic $\delta$ for random positions. We find that 5\% of the random fields have $\delta > 3$.}
        \label{fig:z_random}
    \end{subfigure}
    \hfill
    \begin{subfigure}[t]{0.32\textwidth}
        \centering
        \includegraphics[width=\linewidth]{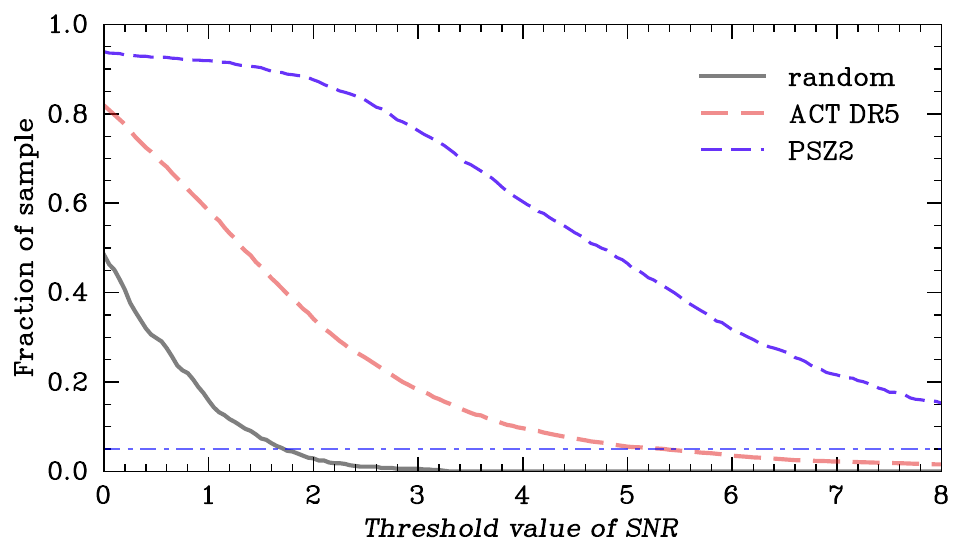}
        \caption{Distribution of SNR values in \texttt{yPlanck} maps. We find that 5\% of random positions have $\mathrm{SNR} > 1.7$.}
        \label{fig:snr_yplanck}
    \end{subfigure}
    \hfill
    \begin{subfigure}[t]{0.32\textwidth}
        \centering
        \includegraphics[width=\linewidth]{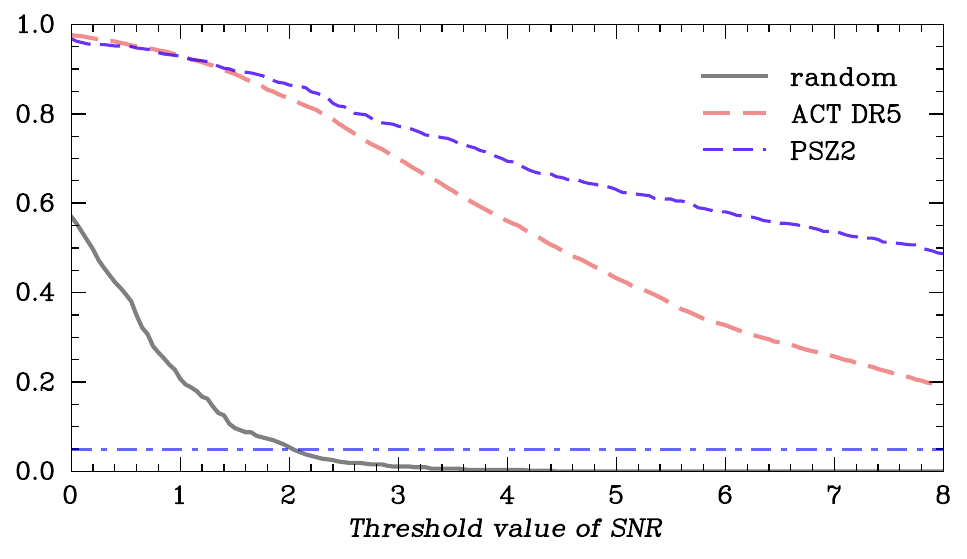}
        \caption{Distribution of SNR values in yACT+\Planck maps. Selection threshold is $\mathrm{SNR} > 2$. All catalogues are restricted  to the ACT DR5 footprint.}
        \label{fig:snr_yact}
    \end{subfigure}
    
    \caption{Fraction of random fields for which the corresponding detection statistic exceeds a given threshold. In each panel, the grey line shows the fraction for random positions. The red dashed line indicates the ACT DR5 candidate sample, while the blue dashed line corresponds to PSZ2. Each panel illustrates the performance of a different detection method: (a) the optical contrast from \texttt{zCluster}; (b) SNR in the yPlanck map; and (c) SNR in the yACT+\Planck\ map.}
    \label{fig:snr}
\end{figure*}

We apply three selection criteria to ensure cluster identification and minimize false detections:

\begin{enumerate}
    \item[(i)] \textbf{\texttt{zCluster} threshold} \label{zCluster_thr} \\
    For the chosen set of templates, we determine a threshold for the optical contrast parameter $\delta$ using a null test on 968 random sky positions. Figure~\ref{fig:z_random} presents the distribution of contrast values for these random fields. We find that 5\% of the random positions yield $\delta > 3$. Based on this analysis, we adopt a contrast threshold that enables the detection of 69 new clusters, while controlling the false detection rate. Lowering the threshold further increases the number of detections but also the contamination rate.

    \item[(ii)] \textbf{\texttt{yACT+Planck} threshold} \label{sec:yact_threshold} \\
    To define the signal-to-noise ratio (SNR) threshold in the \texttt{yACT+Planck} map, we first reproject the map from plate-carée to HEALPix with \texttt{Nside=8192}, preserving the original 0.5 arcmin pixel scale. We then use the same random sample used as in the \texttt{zCluster} null test. Figure~\ref{fig:snr_yact} shows the SNR distribution in random directions (black line). We find that 5\% of random fields exhibit $\mathrm{SNR} > 2$. For comparison, the dashed red line shows the completeness of the ACT DR5 catalogue, which reaches 83.3\% for this SNR threshold. 

    \item[(iii)] \textbf{\texttt{yPlanck} threshold} \label{sec:yplanck_threshold} \\
    For the \texttt{yPlanck} map, we define the SNR as in eq.~\ref{eq:snr}. We apply this estimator to the same set of random fields used above, as well as to known clusters from the PSZ2 catalogue. From this analysis, we determine that 5\% of random positions exceed $\mathrm{SNR} > 1.7$ (see fig.~\ref{fig:snr_yplanck}), which we adopt as the detection threshold for the \texttt{yPlanck}-map sample. With this threshold, the PSZ2 completeness equals to 89 \%.
\end{enumerate}

\section{Mass estimation methodology}
\label{app:mass_estimation}
\begin{figure*}
	\centering
	\begin{subfigure}[b]{0.49\textwidth}
		\includegraphics[width=1\linewidth]{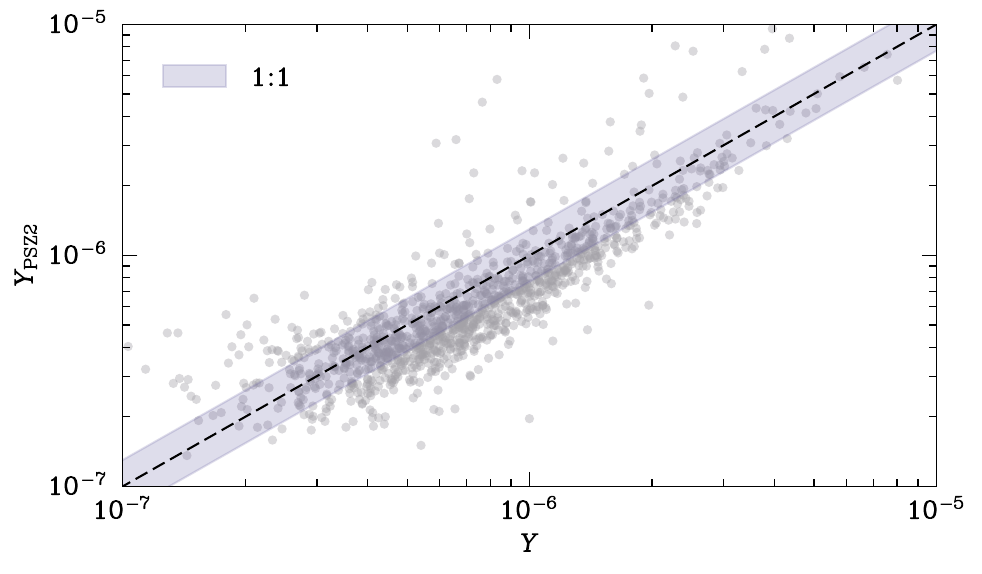}
		\caption{Measurements in the yPlanck map}
		\label{fig:M_pl}
	\end{subfigure}
	\hfill
	\begin{subfigure}[b]{0.49\textwidth}
		\includegraphics[width=1\linewidth]{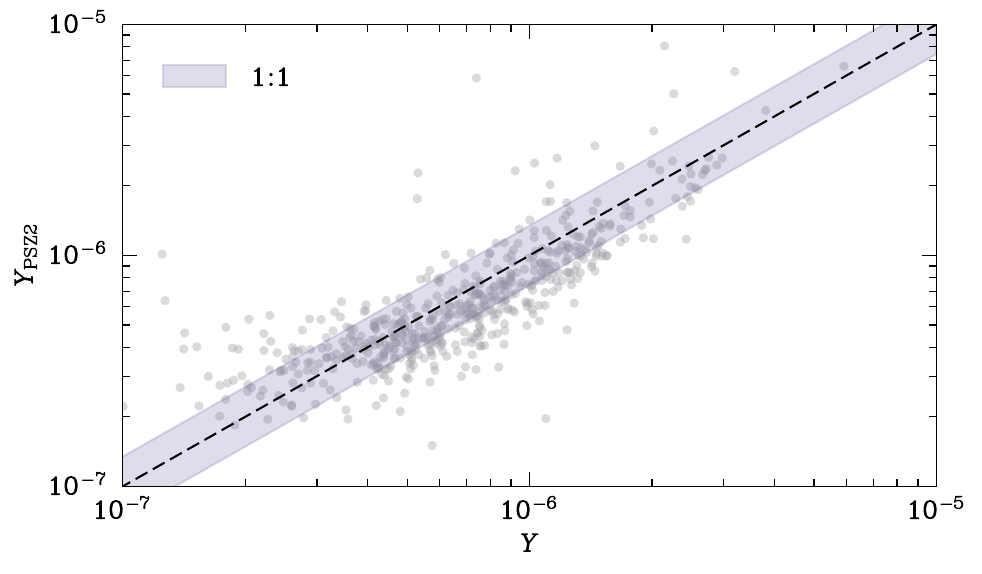}
		\caption{Measurements in the yACT+Planck map}
		\label{fig:M_act}
	\end{subfigure}
	\caption{Dependence of a Compton-$Y$ parameter between ComPACT and PSZ2 catalogue. The purple line represents the scaling relation defined by Equation~\ref{eq:Mpl}. The light purple band shows the $1\sigma$ scatter. Grey points mark individual measurements from the PSZ2 catalogue.}
	\label{fig:mass_combined}
\end{figure*}

Masses for the ComPACT cluster candidates are obtained as follows:
\begin{enumerate}
    \item[(i)] For cluster candidates with available mass estimates (see Table~\ref{tab:catalogs}), we adopt $M_{500c}$ values from the literature. If multiple mass estimates are available for a given cluster (i.e. from SZ, X-ray or optical data), we apply the following priority order: SZ mass measurements, X-ray based mass measurements, and optical masses.Since our mass estimation method is based on SZ measurements and is most directly comparable to \Planck\ masses, SZ-derived masses are prioritized. 
    
    \item[(ii)] For cluster candidates without available mass estimates, we use the  yACT+\Planck\ map to measure masses. 
    First, we extract the "cylindrical" integrated $Y^{\mathrm{cyl}}$ parameter within a 10 arcminute aperture. 
    The following iterative procedure is then applied until convergence is reached:
    \begin{itemize}
        \item We assume an initial value of $R_{500}$ and adopt the conversion coefficient from \cite{Melin_2011} (see their Appendix A), which relates $Y^{cyl}$ to  $Y_{500}$, the SZ flux integrated within a sphere of radius $R_{500}$. To estimate this coefficient, we assume that a gas pressure profile is described by the universal generalized Navarro-Frenk-White (gNFW;  \cite{Nagai_2007})  profile with  concentration parameter $c_{500} = 1.177$,  shape parameters $\alpha = 1.0510$, $\beta = 5.4905$, $\gamma = 0.3081$, and  normalization $P_0 = 8.401$;
        \item The mass $M_{500c}$ is then derived using the scaling relation from \cite{Planck_2014}:
        \begin{equation}
        \label{eq:Mpl}
            E(z)^{-2/3} \left[ \frac{D_A^2 Y_{500}}{10^{-4} Mpc^{2}}\right] = 10^{-0.19} \left[ \frac{(1-b) M_{500c}}{6 \times 10^{14} M_\odot}\right]^{1.79}
        \end{equation}
        
        where $E(z) = \sqrt{\Omega_m (1 + z)^3 + \Omega_\Lambda}$ is the dimensionless Hubble parameter, $D_A$ is the angular diameter distance, and $b = 0.2$ is the hydrostatic bias parameter; 
        \item The procedure is repeated iteratively until convergence in $R_{500}$ is achieved, defined as a fractional  change of less than 10\%.
    \end{itemize}
    
     Since a gNFW profile is assumed, the resulting $Y_{500} - M_{500}$ are not independent of modelling assumptions. We compare our estimates of Compton-Y parameter to those from the PSZ2 catalogue in Figure~\ref{fig:M_act}. The purple line represents the perfect relation (1:1 line), and the shaded region shows the $1\sigma$ scatter. For PSZ2-matched clusters, we find a normalized median absolute deviation (NMAD) for mass estimates of $\sigma = 0.08$ dex. For the ACT DR5 sample, the mass scatter is $\sigma = 0.2$ dex. In addition, although the scaling relation is not explicitly calibrated at high redshift, a comparison with independent ACT DR6 masses shows only a modest increase in scatter from $0.19$ dex at $z < 0.2$ to $0.23$ dex at $z > 0.8$, with no significant redshift-dependent bias ($|\Delta| < 0.05$ dex). This indicates that any systematic uncertainties remain small even at high redshift.

    \item[(iii)] For clusters outside the ACT DR5 footprint, we follow the same methodology as described above but using the yPlanck map. Figure~\ref{fig:M_pl} shows the dependence of the Compton-$Y$ parameter. The NMAD for masses for PSZ2 sample remains $\sigma = 0.08$ dex, but redshift-dependent differences are observed: for $z < 0.1$, the scatter increases to $\sigma = 0.12$ dex, while for $z > 0.1$, it decreases to $\sigma = 0.07$ dex.
\end{enumerate} 

\section{Catalogue simulation}
\label{sec:simul}
In this section, we provide a completeness estimation to illustrate the performance of our catalogue across different mass and redshift ranges. The completeness has been evaluated by injecting simulated clusters into the real ACT+\Planck\ maps and applying the deep learning algorithm to them. We generate parameters for galaxy clusters on a redshift grid that uniformly covers the range $0.2 < z < 2$. The minimum halo mass is set to $M_{500c} > 8 \times 10^{13} M_{\odot}$. In total, we make 1 million samples of simulates clusters. 

For a given redshift and mass, we compute the gas pressure profile describing the electron pressure distribution in the intracluster medium, using the universal gNFW profile \citep{Nagai_2007}:
$$P(\xi) = P_{500} \times \frac{P_0}{(c_{500} \xi)^{\gamma} [(1 + (c_{500}\xi)^\alpha)]^{(\beta - \gamma)/\alpha}}$$

where $P_0 = 8.401$, $c_{500} = 1.177$, $\gamma = 0.3081$, $\alpha = 1.0510$ and $\beta = 5.4905$, as derived in \citet{Arnaud_2010}. The normalization factor $P_{500}$ is computed following the same scaling relations and methodology presented in \citet{Arnaud_2010}.

Next, the pressure profile is interpolated onto a $64 \times 64$ pixel grid in a plate carrée projection and placed at a random position on the ACT+\Planck\ intensity maps. The $64 \times 64$ pixel grid is chosen to ensure that the deep learning model can analyse the cluster surroundings and apply area-based detection criteria. This process is performed separately for three frequencies: 97.8 GHz, 149.8 GHz, and 220 GHz. 

For each cluster, we calculate per-pixel predictions on a $32\times32$ grid centred on the cluster position (because the model input in \citealt{Voskresenskaia_2024} has size $32\times32$ pixels). We apply the minimal threshold of $p_{thr}$ > 0.3 to suppress noise (see \citealt{Voskresenskaia_2024}) and identify the nearest connected group of pixels, characterized by its maximum probability ($p_{max}$) and area ($S$). The detection thresholds are $p_{max} > 0.8$ and $S > 20$ (see figure 5 in \citealt{Voskresenskaia_2024}). 

\end{appendix}

\end{document}